\documentclass[aps,prb,twocolumn,showpacs,10pt]{revtex4-1}

\usepackage[usenames,dvipsnames]{xcolor}

\usepackage{graphicx}
\usepackage{amssymb}
\usepackage{amsmath}
\usepackage{comment}
\usepackage{footmisc}
\usepackage{bbold}
\pdfoutput=1

\begin{document}

\def\pc{\frac{2\pi}{\Phi_0}}

\def\e{\varepsilon}
\def\f{\varphi}
\def\p{\partial}
\def\ba{\mathbf{a}}
\def\bA{\mathbf{A}}
\def\bb{\mathbf{b}}
\def\bB{\mathbf{B}}
\def\bD{\mathbf{D}}
\def\bd{\mathbf{d}}
\def\be{\mathbf{e}}
\def\bE{\mathbf{E}}
\def\bH{\mathbf{H}}
\def\bj{\mathbf{j}}
\def\bk{\mathbf{k}}
\def\bK{\mathbf{K}}
\def\bM{\mathbf{M}}
\def\bm{\mathbf{m}}
\def\bn{\mathbf{n}}
\def\bq{\mathbf{q}}
\def\bp{\mathbf{p}}
\def\bP{\mathbf{P}}
\def\br{\mathbf{r}}
\def\bR{\mathbf{R}}
\def\bS{\mathbf{S}}
\def\bu{\mathbf{u}}
\def\bv{\mathbf{v}}
\def\bV{\mathbf{V}}
\def\bw{\mathbf{w}}
\def\bx{\mathbf{x}}
\def\by{\mathbf{y}}
\def\bz{\mathbf{z}}
\def\bG{\mathbf{G}}
\def\bW{\mathbf{W}}
\def\Bn{\boldsymbol{\nabla}}
\def\Bo{\boldsymbol{\omega}}
\def\Br{\boldsymbol{\rho}}
\def\Bs{\boldsymbol{\hat{\sigma}}}
\def\bh{{\beta\hbar}}
\def\mA{\mathcal{A}}
\def\mB{\mathcal{B}}
\def\mD{\mathcal{D}}
\def\mF{\mathcal{F}}
\def\mG{\mathcal{G}}
\def\mH{\mathcal{H}}
\def\mI{\mathcal{I}}
\def\mL{\mathcal{L}}
\def\mO{\mathcal{O}}
\def\mP{\mathcal{P}}
\def\mT{\mathcal{T}}
\def\mU{\mathcal{U}}
\def\mZ{\mathcal{Z}}
\def\fr{\mathfrak{r}}
\def\ft{\mathfrak{t}}
\newcommand{\rf}[1]{(\ref{#1})}
\newcommand{\al}[1]{\begin{aligned}#1\end{aligned}}
\newcommand{\ar}[2]{\begin{array}{#1}#2\end{array}}
\newcommand{\eq}[1]{\begin{equation}#1\end{equation}}
\newcommand{\bra}[1]{\langle{#1}|}
\newcommand{\ket}[1]{|{#1}\rangle}
\newcommand{\av}[1]{\langle{#1}\rangle}
\newcommand{\AV}[1]{\left\langle{#1}\right\rangle}
\newcommand{\aav}[1]{\langle\langle{#1}\rangle\rangle}
\newcommand{\braket}[2]{\langle{#1}|{#2}\rangle}
\newcommand{\ff}[4]{\parbox{#1mm}{\begin{center}\begin{fmfgraph*}(#2,#3)#4\end{fmfgraph*}\end{center}}}

\def\mr{m_{\perp}}
\def\ml{m_{\parallel}}
\def\hr{H_{\perp}}
\def\hl{H_{\parallel}}

\def\mb{(\mu+\alpha\nu)}
\def\nb{(\nu-\alpha\mu)}
\def\lb{(\lambda+\alpha\kappa)}
\def\kb{(\kappa-\alpha\lambda)}
\def\mn{\left|\bm\times\bz\right|}
\def\etap{\frac{2\pi}{\Phi_0}}
\def\ab{\bar{\alpha}}

\title{Topological Phases of Inhomogeneous Superconductivity}

\author{Silas Hoffman$^1$}
\author{Jelena Klinovaja$^1$}
\author{Daniel Loss$^1$}
\affiliation{$^1$Department of Physics, University of Basel, Klingelbergstrasse 82, CH-4056 Basel, Switzerland}

\pacs{
74.20.-z,%Theories and models of superconducting state
74.25.Ha, % magnetic properties of superconductors
73.63.Nm, %Quantum wires
75.75.-c %Magnetic properties of nanostructures
}

\begin{abstract}
We theoretically consider the effect of a spatially periodic modulation of the superconducting order parameter on the formation of Majorana fermions induced by a one-dimensional system with magnetic impurities brought into close proximity to an $s$-wave superconductor. When the magnetic exchange energy is larger than the inter-impurity electron hopping we model the effective system as a chain of coupled Shiba states. While in the opposite regime, the effective system is accurately described by a quantum wire model. Upon including  a spatially modulated  superconducting pairing, we find, for sufficiently large magnetic exchange energy, the system is able to support a single pair of Majorana fermions with one Majorana fermion on the left end of the system and one on the right end. When the modulation of superconductivity is large compared to the magnetic exchange energy, the Shiba chain returns to a trivially gapped regime while the quantum wire enters a new topological phase capable of supporting two pairs of Majorana fermions. 
\end{abstract}

\maketitle

\section{Introduction}

The interest in topological properties of physical systems has increased following the discovery of topological insulators\cite{pankratovSSC87} and superconductors.\cite{volovikJETP99,hasanRMP10} 
These new phases of matter have garnered much attention for their exotic physical properties and prospective use in classical and quantum computation.\cite{kitaevAoP03,freedmanBotAMS03,bravyiPRA05,bravyiPRA06,aliceaNATP11} In particular, one-dimensional topological superconductors host Majorana fermions (MFs) at their boundaries with nontopological systems which may be used in nearly decoherence-free manipulation of quantum bits.\cite{goldsteinPRB11,budichPRB12,rainisPRB12,schmidtPRB12,zyuzinPRL13,pedrocchiPRL15}

Recent theoretical\cite{choyPRB11,nadj-pergePRB13,pientkaPRB13,klinovajaPRL13,vazifehPRL13,brauneckerPRL13} and experimental\cite{nadj-pergeSCI14,pawlakCM15} efforts have been made to engineer such MFs by placing magnetic adatoms on the surface of a superconductor. When the magnetic exchange between an adatom and the quasiparticles in the bulk superconductor is sufficiently strong, a local Shiba state\cite{yuAPS65,shibaPTP68,rusinovJETP69b} with energy inside the superconducting gap is formed. A chain of such adatoms creates many Shiba states that can hybridize and form a band, which can itself support zero energy MFs at the ends of the chain.\cite{pientkaPRB13,nadj-pergeSCI14} Alternatively, when the magnetic exchange energy is small compared to the inter-adatom coupling, a one-dimensional band with helical magnetic order, via the Ruderman-Kittel-Kasuya-Yosida interaction,\cite{klinovajaPRL13,vazifehPRL13,brauneckerPRL13} is formed in the adatom chain itself that, analogous to a one-dimensional quantum wire with proximity-induced superconductivity,\cite{brauneckerPRB10,kjaergaardPRB12,klinovajaPRB12} can form MFs at the ends of the chain.\cite{pawlakCM15}

It is known that magnetic impurities within superconductors cause a local decrease in the superconducting order parameter.\cite{rusinovJETP69,schlottmanPRB76,salkolaPRB97,balatskyRMP06,flattePRL97,flattePRB97,flatteSSP99,mengPRB15,hoffmanPRB15,bjornsonCM15} When the exchange energy is of the order the Fermi energy, the pairing potential can be reduced to zero and even reverse sign,\cite{flattePRL97,salkolaPRB97} forming a local $\pi$-junction. Away from the impurity, the superconducting gap returns to the bulk value described by a power-law dependence with the lengthscale set by the Fermi wavelength. Consequently, a chain of such impurities induces a periodic spatial modulation of the pairing potential.  The effect of such modulation of the superconductivity strength on the formation of MFs in a chain of spin impurities has not been studied. In this paper, we address this in both previously mentioned regimes. When a band of Shiba states is formed and the interatomic electron hopping can be neglected,\cite{pientkaPRB13,pientkaPRB14} we find two phases as a function of the magnetic exchange energy and periodic decrease in pairing potential: one in which there are no MFs and one in which there is a single pair of MFs with one on the left end of the system and one on the right end. Conversely, when the interatomic electron hopping is larger than the magnetic exchange energy, the system can support three phases: in addition to the known trivial phase with no MFs and the topological phase supporting a single pair of MFs,\cite{klinovajaPRB12} for a sufficiently large amplitude in the inhomogeneity of the order parameter, two pairs of MFs can be realized with two (orthogonal) MFs on the left end of the system and two on the right end.

\begin{figure}
\includegraphics[width=1\linewidth]{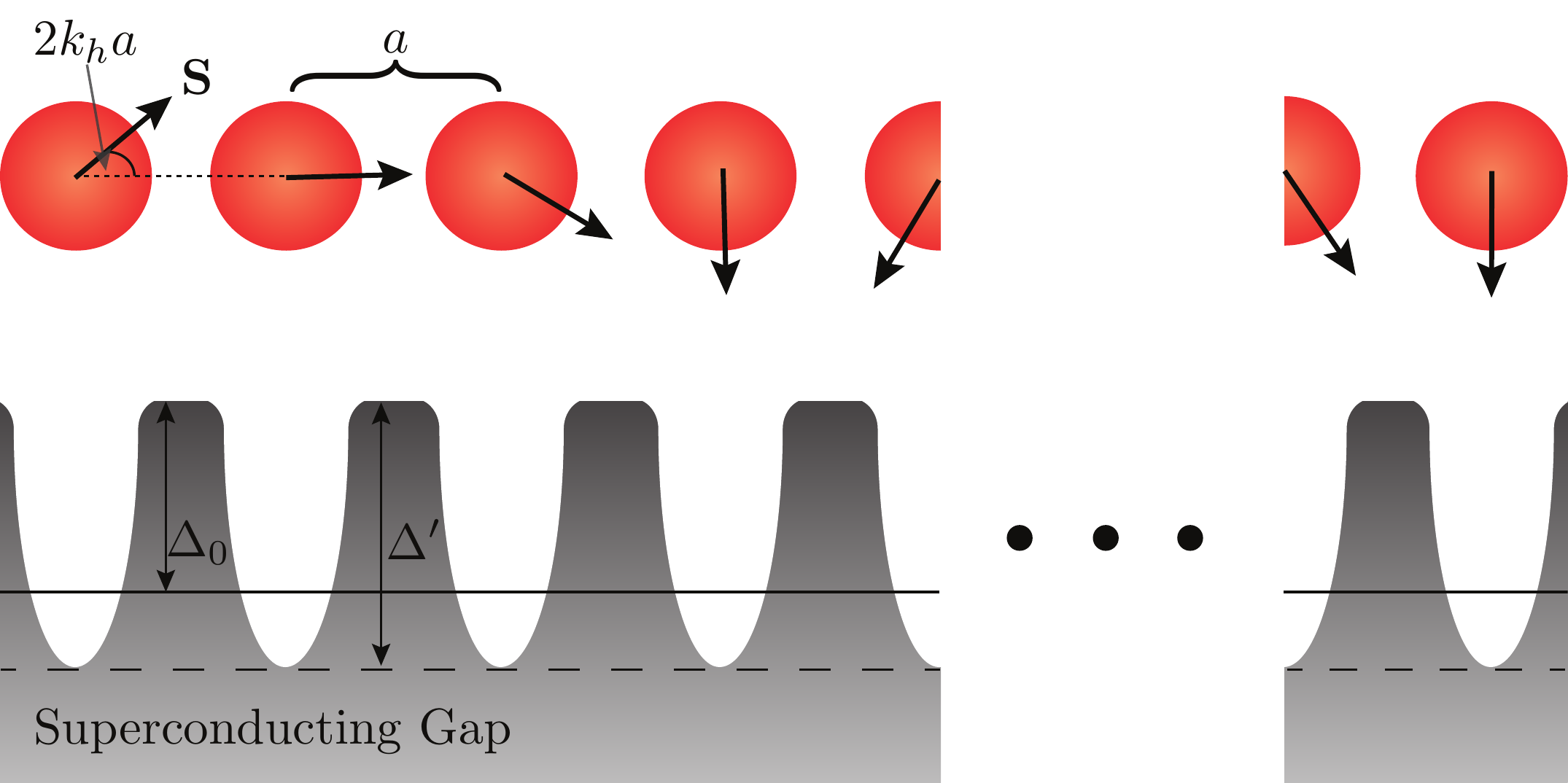}
\caption{Schematic of our setup of a chain of helically ordered magnetic impurities with pitch $2k_ha$ and magnetic dipole moment $\textbf S$ on top of a conventional $s$-wave superconductor with bulk gap $\Delta_0$, where $a$ is the distance between adjacent impurities. At the point of each impurity, the superconducting order parameter is locally reduced by $\Delta'$. 
}
\label{setup}
\end{figure}

The paper is organized as follows. In Sec. II, we detail the Shiba chain model with periodic decrease in superconductivity that we use to find the MF bound state solutions, allowing us to qualitatively describe the phase diagram as a function of this decrease in superconductivity and magnetic exchange energy. Using the model of a quantum wire in Sec.~\ref{qw}, we again construct the phase diagram, as a function of the same control parameters by explicitly finding the MF wavefunctions. We conclude in the final section, Sec. IV, by considering the implication on current experimental realizations of magnetic impurity chains and suggest systems where these effects would be most pronounced.

\section{Shiba Chain Model}
\label{shiba}
We begin with the case when the inter-impurity coupling strength is small compared to the magnetic exchange strength with the quasiparticles in the superconductor. However, before we model a \textit{chain} of magnetic impurities, we recall the properties of a \textit{single} magnetic impurity within a conventional $s$-wave superconductor. We consider the Hamiltonian\cite{mengPRB15}
 \eq{
 H=\xi_\textbf{p}\tau_z+\Delta(\textbf r)\tau_x - J\textbf{S}\cdot\boldsymbol\sigma\delta(\textbf{r})\,,
\label{BdG}
}
where $\xi_\textbf{p}=\textbf p^2/2m-\mu$ is the dispersion of the quasiparticles with momentum $\bf p$ and mass $m$ in the normal metal phase, $\mu$ is the chemical potential, and $\Delta(\textbf r)$ is the local superconducting pairing potential at position $\textbf{r}$. The Pauli matrices $\boldsymbol \tau$ ($\boldsymbol\sigma$) act in Nambu (spin) space in the basis $(\Psi_\uparrow,\Psi_\downarrow,\Psi_\downarrow^\dagger,-\Psi_\uparrow^\dagger)$ where $\Psi_\sigma$ ($\Psi_\sigma^\dagger$) are the electron creation (annihilation) operators. Here, $J$ is the magnetic exchange interaction strength between the magnetic impurity  and the quasiparticles; $\textbf S=(\cos\phi\sin\theta,\sin\phi\sin\theta,\cos\theta)$ is the magnetic moment of the impurity, treated as classical variable. We consider a homogenous pairing potential, $\Delta_0$ with a local decrease at the position of the impurities given by $\Delta(\textbf{r})=\Delta_0- l^3\Delta'\delta(\textbf r)$ that models a reduction in the gap by $\Delta'$ in the volume $l^3$ around the impurity.

Defining $\alpha = \pi \nu_F J S$ and $\alpha'=\pi\nu_Fl^3 \Delta'$, where $\nu_F$ is the density of states at the Fermi energy and $S=|\textbf S|$, one may show that, when $\alpha<\alpha'$, there are four in-gap bound state solutions with wavefunction at the position of the impurity ${\bf r} =0$:
\begin{align}
\psi_+&=\left(\begin{array}{c}|\uparrow\rangle \\|\uparrow\rangle
\end{array}\right)\,,\,\,
\psi_-=-\left(\begin{array}{c} |\downarrow\rangle\\|\downarrow\rangle
\end{array}\right)\,,\,\,\nonumber\\
\chi_+&=\left(\begin{array}{c}-|\uparrow\rangle\\|\uparrow\rangle
\end{array}\right)\,,\,\,
\chi_-=\left(\begin{array}{c}  |\downarrow\rangle\\- |\downarrow\rangle
\end{array}\right)\,,\,\,
\label{basis}
\end{align}
where
\begin{align}
 |\uparrow\rangle = \left(\begin{array}{c}\cos\theta/2\\e^{i\phi}\sin\theta/2
\end{array}\right)\,,\,\,
 |\downarrow\rangle = \left(\begin{array}{c}e^{-i\phi}\sin\theta/2\\-\cos\theta/2
\end{array}\right)\,.
\label{spin}
\end{align}
These bound states have energies $E_{+,+}$, $E_{+,-}$, $E_{-,+}$, and $E_{-,-}$, respectively, where
\begin{align}
\frac{E_{\sigma,\tau}}{\Delta_0}=\tau\frac{ 1-\beta_{\sigma,\tau}^2}{1+\beta_{\sigma,\tau}^2}\,.
\end{align}
Here, $\tau$ is the eigenvalue of $\tau_x$, $\sigma$ is the eigenvalue of $\textbf{S}\cdot\boldsymbol\sigma$, and $\beta_{\sigma,\tau}=\tau\alpha'+\sigma\alpha$.
When  $\alpha>\alpha'$, the magnetic exchange energy overcomes the decrease in superconducting order parameter and there are only two solutions, $\psi_+$ and $\chi_-$, to Eq.~(\ref{BdG}).

Next, to model a chain of such impurities (see Fig.~\ref{setup}), we consider the Hamiltonian 
\eq{
 H=\xi_\textbf{p}\tau_z+\Delta(\textbf r)\tau_x -\sum_{i=1}^N J_i\textbf{S}_i\cdot\boldsymbol\sigma\delta(\textbf{r}_i-\textbf{r})\,,
\label{BdG}
}
where $\textbf{r}_i$ are the positions of the $N$ impurities and take the local decrease in superconductivity as $\Delta(\textbf{r})=\Delta_0- l^3\sum_{i=1}^N\Delta'_i\delta(\textbf r_i)$. In the following we consider identical impurities, so that $J_i=J$, $|\textbf{S}_i|=S$, and $\Delta_i'=\Delta'$. We  consider a planar helix, $\theta_i=\pi/2$, with pitch $k_h$, $\phi_i=2k_h r_i$, where $\theta_i$ and $\phi_i$ parameterize the magnetic dipole moment at $\textbf r_i$ according to Eq.~(\ref{spin}).

The four-component Bogoliubov-deGennes (BdG) spinor $\psi(\textbf{r}_i)$ with energy $E$ at $\textbf{r}_i$ satisfies the coupled equations which we find following the procedure of Ref.~[\onlinecite{pientkaPRB13}]
\begin{align}
\psi(\textbf r_i)&=(\alpha\hat J_i \textbf s_i\cdot\boldsymbol\sigma+\alpha'\hat J_i \tau_x)\psi(\textbf r_i)\nonumber\\
&+\sum_{j\neq i}(\alpha\hat \Gamma_{ij}\textbf s_j\cdot\boldsymbol\sigma+\alpha'\hat\Gamma_{ij}\tau_x)\psi(\textbf r_j)\,,
\label{EoM1}
\end{align}
where $\textbf s_i=\textbf S_i/S_i$ and
\begin{align}
&\hat J_i=\frac{ 
E +\tau_x \Delta_0 }{\sqrt{\Delta_0^2-E^2}}\,,\nonumber\\
&\hat\Gamma_{ij}=\left[\frac{ (E+\tau_x \Delta_0  )\sin (k_F r_{ij})} {\sqrt{\Delta_0^2-E^2}}+\tau_z\cos (k_Fr_{ij}) \right]
\frac{e^{-r_{ij}/\xi_E}}{k_F r_{ij}}\,.
\label{EoM2}
\end{align} 
The distance between impurities at $\textbf r_i$ and $\textbf r_j$ is $r_{ij}=|\textbf r_i-\textbf r_j|$, $k_F$ ($v_F$) is the Fermi wave vector (Fermi velocity) of the superconductor in the normal metal state and  $\xi_E=\hbar v_F/\sqrt{\Delta_0^2-E^2}$. We only consider the regimes when the bound states are close to the chemical potential, $\beta_{\sigma,\tau}\approx\pm1$, so that the uncoupled states have energy $E_{\sigma,\tau}\approx\tau\Delta_0(1-|\beta_{\sigma,\tau}|)$. Further, we expand to lowest order in the coupling between impurity sites $1/k_Fr$. 

There are two regimes of the Shiba band model [see Eq.~(\ref{EoM1})] accessible in this approximation. When the magnetic exchange energy is greater than the decrease in superconductivity ($\alpha>\alpha'$ but not necessarily $\alpha\gg\alpha'$), there are only  two bands in the spectrum and, because $\beta_{+,+}=\beta_{-,-}\approx1$, both are near the chemical potential. When the magnetic exchange energy is much smaller than the decrease in superconductivity ($\alpha\ll\alpha'\approx1$), there are four bands in the spectrum: two slightly above the chemical potential at energy with energy difference $2\alpha$, $E_{\pm,+}=\Delta_0(1-\alpha'\mp\alpha)$, and two at $-E_{\pm,+}$. We exclude the regime $\alpha<\alpha'$, because, although two of the particle-hole partners will have a  energy near the chemical potential, the other pair will have energy somewhere between the chemical potential and the gap edge and thus are not captured by our approximation. \footnote{We expect if $\alpha\approx\alpha'$ and the overlap of Shiba states is small, so that the up and down spin are nearly decoupled, we can project out the higher energy bands to obtain an effective two-band model with energies near the chemical potential, similar to the $\alpha>\alpha'$ regime.} 

When $\alpha>\alpha'$, while keeping $\alpha+\alpha'\approx\pm1$, there are only two bands within the gap. We construct an effective Hamiltonian by projecting Eq.~(\ref{EoM1}) onto the unperturbed eingenspinors of Eq.~(\ref{basis}), $\psi_+$ and $\chi_-$, to obtain a $2N\times2N$ matrix equation, $H^{\textrm{eff}}_1\phi_1=E\phi_1$ for an $N$ impurity chain written in the basis $(\psi_{+i},\chi_{-i})$, where (see Appendix~\ref{idS} for some useful identities)
\begin{align}
H_1^{\textrm{eff}}&=\left(\begin{array}{cc}
 h^{\textrm{eff}}_{ij} & \Delta^{\textrm{eff}}_{ij}\\
 (\Delta^{\textrm{eff}}_{ij})^\dagger & - h^{\textrm{eff}}_{ij}
\end{array}\right)\,,
\label{heff1}\nonumber\\
h^{\textrm{eff}}_{ii}&=E_{+,+}\,,\,\,\,\,\,\,\,\,\,\Delta^{\textrm{eff}}_{ii}=0\,,\nonumber\\
h^{\textrm{eff}}_{ij}&=-\Delta_0\frac{e^{-r_{ij}/\xi_0}}{k_F r_{ij}}\sin(k_Fr_{ij})\langle\uparrow,i|\uparrow,j\rangle\,,\nonumber\\
\Delta^\textrm{eff}_{ij}&=\Delta_0\frac{e^{-r_{ij}/\xi_0}}{k_F r_{ij}}\cos(k_F r_{ij})\langle\uparrow,i|\downarrow,j\rangle\,,
\end{align}
for $i\neq j$. Here, $\psi_{+i}$ and $\chi_{-i}$ ($|\uparrow,i\rangle$ and $|\downarrow,i\rangle$) are the spinors defined in Eq.~(\ref{basis}) [in Eq.~(\ref{spin})] at $\textbf r_i$. After taking the Fourier transform (see Appendix~\ref{aft}) and making a unitary rotation by $\pi/2$ around the $x$ axis in Nambu space, the Hamiltonian takes the off-diagonal form
\eq{
\mathcal H_1^{\textrm{eff}}=h_1\tau_y+\Delta_1\tau_x\,,
}
with eigenvalues and eigenvectors $\epsilon_\pm=\pm\sqrt{h_1^2+\Delta_1^2}$ and $\left(h_1+\epsilon_\pm,~\Delta_1\right)^T $, respectively, where
\begin{align}
h_1&=\frac{\Delta_0}{k_F a}\Im\left\{\ln\left[f(k_F+k_+)f(k_F-k_+)\right.\right.\nonumber\\
&\left.\left.\times f(k_F+k_-)f(k_F-k_-)\right]\right\}+\Delta_0(1-\alpha-\alpha')\,,\nonumber\\
\Delta_1&=-\frac{\Delta_0}{k_F a}\Re\left\{\ln\left[\frac{f(k_F+k_+)f(k_F-k_+)}{f(k_F+k_-)f(k_F-k_-)}\right]\right\}\,.
\label{ham12}
\end{align}
 $f(k)=1-\exp(-a/\xi_0+i a k)$, $k_\pm=k\pm k_h$, and $a$ is the impurity spacing.\footnote{Because this is the logarithm of a complex function, we should specify the Riemann sheet for definiteness. However, because this shifts $\kappa_i^\pm$ only by $2\pi i n$, the MF solution is unchanged by this choice.}   Here, $\Re[F(k_F,k_\pm)] = [F(k_F,k_\pm)+F(-k_F,-k_\pm)]/2$ and $\Im[F(k_F,k_\pm) ]= [F(k_F,k_\pm)-F(-k_F,-k_\pm)]/2i$ which reduce to the real and imaginary parts, respectively, when $k_\pm$ is real. The zero energy solutions are defined by $h_1=\pm i\Delta_1$. In order to find the MF solution, we linearize Eq.~(\ref{ham12}) around the Dirac points in the continuum limit $1/a\gg k_F,~k_h$. When $k_F>k_h$ there are four Dirac points,
 \begin{align}
 k_1&=0\,,\,\,\,k_2=\pi/a\,,\nonumber\\
 k_3&=-k_4=-\sqrt{k_F^2-k_h^2}\,.
 \end{align}
When $k_F<k_h$, there are only two Dirac points, $k_1$ and $k_2$. We Fourier transform the linearized Hamiltonian back into real space and find the zero energy eigenvectors are $\Phi^\pm_\mu (r)=(\pm i,~1)^T e^{i k_\mu r - \kappa_\mu^\pm r}$, where $r=|\textbf r_i - \textbf r_0|=r_{i0}$ is the distance away from the left end written as a continuous variable, and
\begin{align}
\kappa_1^\pm&=\pm \pi\frac{k_F^2-k_h^2}{2k_h}\,,\,\,\,\kappa_2^\pm=\pm (3-\alpha-\alpha')\frac{k_F}{k_h a}\,,\nonumber\\
\kappa_3^\pm&=\kappa_4^\pm=\mp\pi k_h/2\,.
\label{decay}
\end{align}
When $k_F>k_h$ $\mu=1,\ldots,4$, while $\mu=1,2$ when $k_F<k_h$, as there are only two Dirac cones. 

In order for a MF to form at the left end of the chain, $r=0$, the wave function must vanish at both infinity, 
$\Phi^M(r\rightarrow\infty)=0$, and the left end, $\Phi^M(0)=0$. To satisfy the former boundary condition, one must choose the values of $\kappa_\mu^\pm$ in Eq.~(\ref{decay}) corresponding to exponentially decaying solution. Therefore, because $\Phi^+_\mu$ and $\Phi^-_\nu$ are orthogonal, the latter condition can be satisfied if and only if two different $\kappa_\mu^+$ or $\kappa_\mu^-$ have the same sign, {\it i.e.}, $\kappa^\pm_\mu \kappa^\pm_\nu>0$ for $\mu\neq\nu$, which is true only when $0<k_h<k_F$. In this case, we find the Majorana wave function, up to normalization, to be
\eq{
\Phi^M(r)= e^{i\pi/4}[\Phi^-_3(r)-\Phi^-_4(r)]\,,
}
in agreement with Ref.~[\onlinecite{pientkaPRB14}] in the special case that $\alpha'\rightarrow0$. 
One finds a Majorana wave function on the opposite side of the chain, using an analogous argument, taking $\Phi_\mu^-\rightarrow\Phi_\mu^+$.

When the decrease in the local superconductivity is much larger than the magnetic exchange energy, $\alpha'\gg\alpha$, while keeping all uncoupled Shiba energies near the chemical potential, $E_{\sigma,\tau}\approx0$, one must use all four solutions of the uncoupled impurity [Eq.~(\ref{basis})]. Following the same procedure as the two band model, one may show that there are not MFs possible in this parameter regime. See Appendix~\ref{asb} for technical details.

\begin{figure}
\includegraphics[width=1\linewidth]{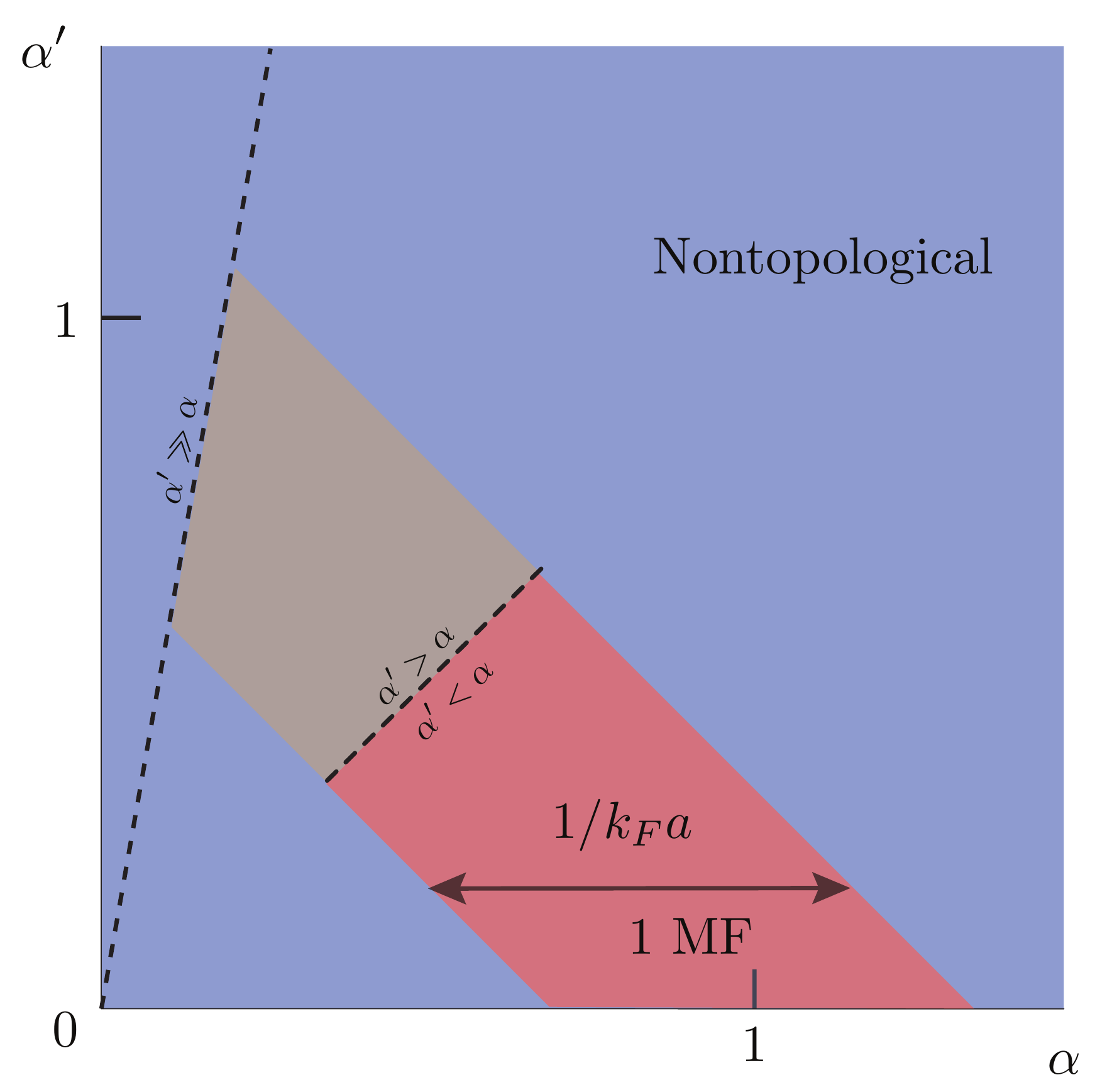}
\caption{Qualitative phase diagram for the helical impurity chain as modeled by a Shiba band within the bulk superconductor. The system is able to support a single pair of MFs if the Shiba band is near the chemical potential (within  the bandwidth), $\alpha+\alpha'\approx1$, the decrease in superconductivity is smaller than the magnetic exchange energy, $\alpha'\lesssim\alpha$, and the pitch of the helix is larger than the Fermi wavelength, $0<k_h<k_F$.  The grey region  cannot be described by our model. Otherwise, the system is topologically trivial.}
\label{phase1}
\end{figure}

Collecting our results in a qualitative phase diagram (Fig.~\ref{phase1}), we find that for a hybridized Shiba band to be in a topologically nontrivial phase the bands must satisfy three conditions. First, the uncoupled Shiba states must be approximately within the width of the band ($\sim\Delta_0/k_Fa$) to the chemical potential, so that some of the bands are filled and some are empty. Second, the magnetic exchange energy must be large enough compared to the reduction in superconductivity, $\alpha\gtrsim\alpha'$, so that there are two bands near the chemical potential. Lastly, $k_F>k_h$.

\section{Quantum Wire Model}
\label{qw}
In this section, we consider the regime where the magnetic exchange interaction between the spins of the electrons of the bulk superconductor with the magnetic atoms forming the chain is much smaller than the intrachain electron hopping. The superconductor induces a proximity gap in the chain of magnetic atoms which, in the ground state, are helically ordered with pitch twice the Fermi wave vector, $2k_F$,\cite{klinovajaPRL13,vazifehPRL13,brauneckerPRL13} where the Fermi wavelength is now defined by the Fermi level of the wire. As such, we model the  chain as a quantum wire with a helical magnetic order on top of a superconductor by
\begin{align}
H^0=&H^{\textrm{kin}}+H^Z+H^{sc}\,.
\label{Hf}
\end{align}
The contributions to the Hamiltonian from the kinetic energy, magnetic exchange interaction, and proximity induced superconductivity are given by
\begin{align}
&H^\textrm{kin}=\sum_{\sigma} \int dr\  \Psi_\sigma^\dagger (r)\left[\frac{(-i \hbar \partial_r)^2}{2m}-\mu\right]\Psi_{\sigma} (r)\,, \nonumber\\
&H^Z=\Delta_z \sum_{\sigma, \sigma'} \int dr \ \Psi_\sigma^\dagger (r)\large[\cos(2k_F r)\sigma_x\nonumber\\
&\hspace{57pt}+\sin(2k_Fr)\sigma_y\large]_{\sigma \sigma'}\Psi_{\sigma'} (r)\,,\nonumber\\
&H^{sc}=\frac{\Delta_{0}}{2}\sum_{\sigma, \sigma'}\int dr\ [ \Psi_\sigma(r)(i\sigma_y)_{\sigma\sigma'} \Psi_{\sigma'}(r)+\textrm{h.c.}]\,, \nonumber\\
\end{align}
respectively, where $m$ is the effective electron mass, $\mu$ is the chemical potential of the wire, $\Delta_0$ is the uniform proximity induced superconducting order parameter, and $\Delta_z$ is the splitting energy of two opposite spin directions (as a result of the magnetic exchange interaction). The operator $\Psi_\sigma$ ($\Psi^\dagger_\sigma$)  destroys (creates) an electron with spin $\sigma=\uparrow,\downarrow$ and $\sigma_i$ are the Pauli matrices acting on spin space. The Hamiltonian can be linearized around $\pm k_F$ by expressing the electron operators in terms of slowly varying left and right movers,\cite{brauneckerPRB10,klinovajaPRB12,klinovajaPRL13}  $\Psi_\sigma(r)=R_\sigma(r)e^{ik_Fr}+L_\sigma(r)e^{-ik_Fr}$ to obtain two modes, $\psi_r=(R_\uparrow,L_\downarrow,R_\uparrow^\dagger,L_\downarrow^\dagger)$ and $\psi_l=(L_\uparrow,R_\downarrow,L_\uparrow^\dagger,R_\downarrow^\dagger)$, of opposite helicity, where $\psi_r$ and $\psi_l$ are written in Nambu space. The components of the Hamiltonian density, $\mathcal H_0$, defined by
\begin{equation}
H^0=\frac{1}{2}\int dr \psi^\dagger \mathcal H^0 \psi =  \frac{1}{2}\int dr \psi^\dagger (\mathcal H^{\textrm{kin}}+ \mathcal H^{\textrm{sc}}+ \mathcal H^{\textrm{Z}})\psi\,,
\end{equation} reduce to
\begin{align}
\mathcal H^{\textrm{kin}}&=(-i\hbar v_F\partial_r)\sigma_z\otimes\mathbb1\otimes\rho_z\,,\nonumber\\
\mathcal H^{Z}&=\Delta_z\sigma_x\otimes\eta_z\otimes(1+\rho_z)/2\,,\nonumber\\
\mathcal H^{sc}&=\Delta_0\sigma_y\otimes\eta_y\otimes\mathbb1\,,
\end{align}
after dropping all fast oscillating modes where $\psi=(\psi_r,\psi_l)$ and $v_F$ is the Fermi velocity. Here, $\eta_i$ and $\rho_i$ are the Pauli matrices acting in Nambu space and mode space (\textit{e.g.}, $\rho_x$ exchanges $\psi_r$ and $\psi_l$), respectively; $\psi_r$ and $\psi_l$ are uncoupled by $\mathcal H_0$.

If the superconducting order parameter is not uniform, we include a spatial modulation of $\tilde\Delta$. One may show that after dropping fast oscillating modes, the only additional contribution to the Hamiltonian, $H^I=(1/2)\int dr\ \psi^\dagger \mathcal H^I \psi$, is modulation of the superconducting order with wavelength $\lambda_F/2$,
\eq{
\mathcal H^I=\Delta'\sigma_y\otimes\eta_y\otimes\rho_x\,,
\label{Hi}
}
where $\Delta'$ is the amplitude of the Fourier mode of $\tilde\Delta$ at $2k_F$; $\mathcal H^I$ mixes the modes $\psi_r$ and $\psi_l$. 

The bulk spectrum of the full Hamiltonian, $H=H^0+H^I$,\footnote{One may show the Hamiltonian $H$ is invariant under the the (pseudo)time-reversal operation $\mathcal T$ and particle-hole symmetry $\mathcal{C}$ defined by $U_{T}^\dagger (H^*) U_{T}=H$ and $U_{C}^\dagger (H^*) U_{C}=-H$, respectively. Because the matrices $U_{T}=\sigma_x\otimes\eta_z\otimes\mathbb1$ and $U_\mathcal{C}=\sigma_z\otimes\mathbb1\otimes\mathbb1$ satisfy the property $U_T^*U_T=U_C^*U_C$, this Hamiltonian is the BDI topological class\cite{ryuNJoP10} of Hamiltonians.} is given by
\begin{align}
&\mathcal E^2_{\sigma\sigma'}=(\hbar v_Fk)^2+\Delta_0^2+\Delta'^2+(\Delta_0+\sigma\Delta_z/2)^2+(\Delta_z/2)^2\nonumber\\
&+4\sigma'\sqrt{(\hbar v_F k\Delta')^2+(\Delta_0+\sigma\Delta_z/2)^2(\Delta'^2+\Delta^2/4)}\,.
\label{energy}
\end{align}
The eight eigenstates of $H$ with zero energy, $\mathcal E_{\sigma\sigma'}=0$, after transforming to the electron-hole basis, $(\Psi_\uparrow,\Psi_\downarrow,\Psi^\dagger_\uparrow,\Psi^\dagger_\downarrow)$, and dropping fast oscillating terms, can be written as $\Phi_i^\pm(r)=\chi_i^\pm\exp(-\kappa_i^\pm r)$ (see Appendix~\ref{aqw}) where $i=1\dots4$ and 
\begin{align}
\hbar v_F\kappa^\pm_1&=-\frac{1}{2}\left[\Delta_z\pm\sqrt{(\Delta_z-2\Delta_0)^2-4\Delta'^2}\right]\,,\nonumber\\
\hbar v_F\kappa^\pm_2&=-\frac{1}{2}\left[-\Delta_z\pm\sqrt{(\Delta_z+2\Delta_0)^2-4\Delta'^2}\right]\,,\nonumber\\
\hbar v_F\kappa^\pm_3&=-\frac{1}{2}\left[-\Delta_z\pm\sqrt{(\Delta_z-2\Delta_0)^2-4\Delta'^2}\right]\,,\nonumber\\
\hbar v_F\kappa^\pm_4&=-\frac{1}{2}\left[\Delta_z\pm\sqrt{(\Delta_z+2\Delta_0)^2-4\Delta'^2}\right]\,.\nonumber\\
\label{ev1}
\end{align}

As in the previous section, our goal is to construct a MF wave function that satisfies the boundary conditions. Our strategy is to identify the zero energy solutions that decay exponentially with increasing $r$; the sign of the real part of the wave vectors in Eq.~(\ref{ev1}) changes as $\kappa_i^\pm$ goes through zero as a function of the parameters $\Delta_z$, $\Delta_0$, and $\Delta'$. One may see from Eq.~(\ref{ev1}) that this happens when $\Delta_z=|\Delta_0^2-\Delta'|^2/\Delta_0$ wherein the gap closes and there is a  change in the topological phase.

When $0\leq\Delta_z<(\Delta_0^2-\Delta'^2)/\Delta_0$, the solutions that decay exponentially are 
$\Phi^-_1,~\Phi^-_2,~\Phi^-_3$, and $\Phi^-_4$. 
However, because $\Phi_i^\pm(r)$ are mutually orthogonal for any value of $r$, one cannot form a MF that satisfies the boundary condition at the wire end and therefore this phase defined by that boundary is trivial.

If $\Delta_z>|\Delta_0^2-\Delta'^2|/\Delta_0$, the decaying solutions are $\Phi_2^-,~\Phi_3^+,~\Phi_3^-$, and $\Phi^-_4$. We find one MF wave function satisfying the boundary conditions:
\begin{align}
\Phi^M_1(r)=&i[(\hbar v_F\kappa_3^-+\Delta_0-\Delta')\Phi_3^+(r)\nonumber\\
&-(\hbar v_F\kappa_3^++\Delta_0-\Delta')\Phi_3^-(r)]\,.
\end{align}
Therefore, this region is topologically nontrivial.  When $\Delta'=0$, the solution and topological criteria ($\Delta_z>\Delta_0$) reduces to the result found in Ref.~[\onlinecite{klinovajaPRB12}].

When $0<\Delta_z<(\Delta'^2-\Delta_0^2)/\Delta_0$, the decaying solutions are $\Phi_2^+,~\Phi_2^-,~\Phi_3^+$, and $\Phi_3^-$. In this regime $\Delta'>|\Delta_0\pm\Delta_z|$, so that $\kappa_i$ acquire an imaginary part and the decay length of the MF solutions is set by $1/\Delta_z$. We find, in addition to the MF already constructed, $\Phi_1^M(r)$, another MF at the same end
\begin{align}\Phi^M_2(r)=&i[(\hbar v_F\kappa_2^-+\Delta_0-\Delta')\Phi_2^+(r)\nonumber\\
&-(\hbar v_F\kappa_2^++\Delta_0-\Delta')\Phi_2^-(r)]\,.
\end{align}
If $0=\Delta_z<(\Delta'^2-\Delta_0^2)/\Delta_0$ we may still construct two MF solutions satisfying the boundary conditions but they are delocalized, \textit{i.e.}, the decay length is infinite. Although $\Phi^M_1$ and $\Phi^M_2$ are orthogonal, there is no inherent symmetry in the system that prevents two MFs at the same end from splitting, \textit{e.g.}, they are not Kramer's pairs. One may show that an infinitesimal deviation away from a planar helix will split the MF energies away from zero. If this deviation is small, which is expected for systems with large anisotropy, the splitting is negligible and the energy of the two pairs of MFs remains within the gap.

\begin{figure}
\includegraphics[width=1\linewidth]{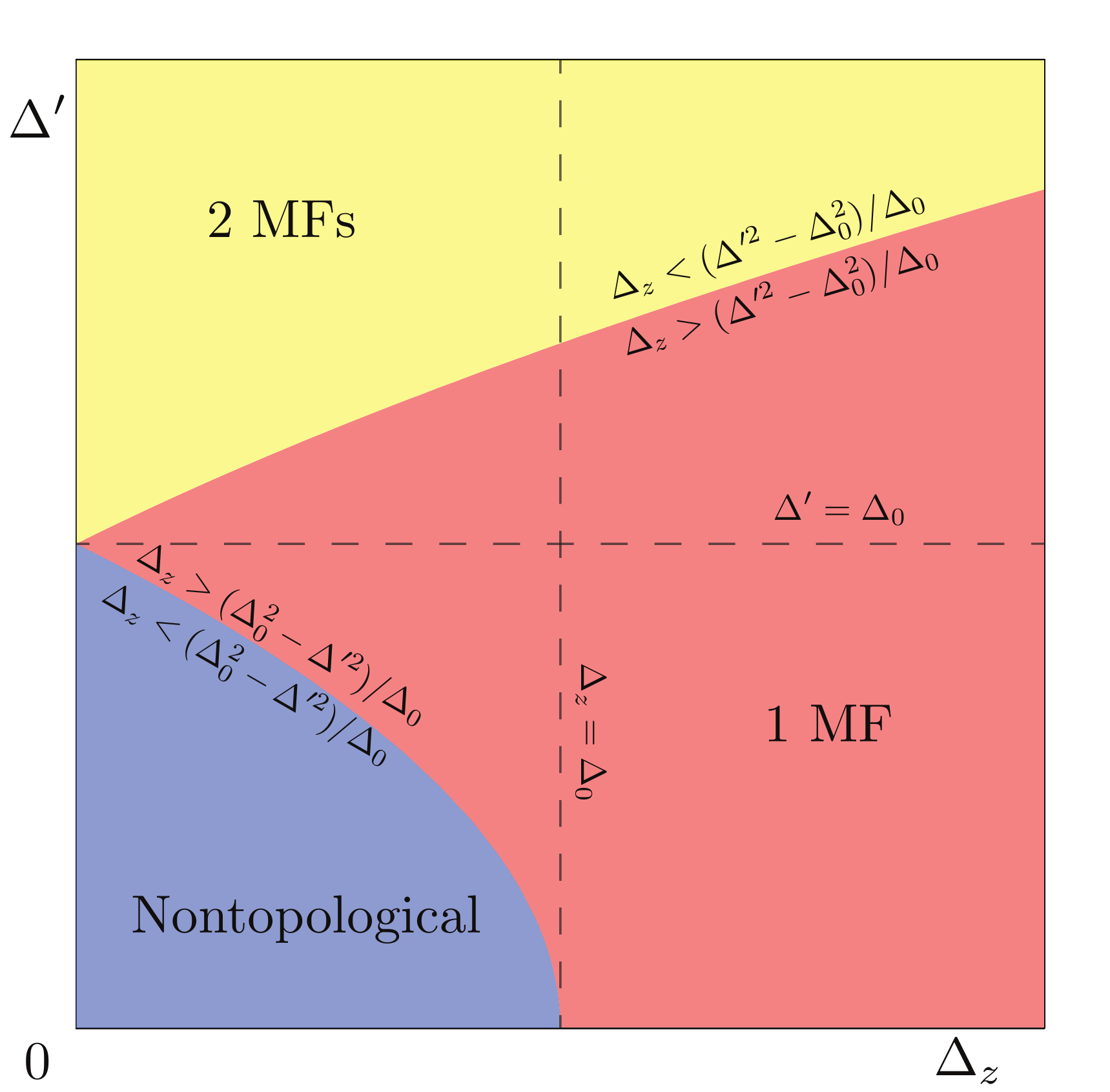}
\caption{Qualitative phase diagram for the helical impurity chain as modeled by a quantum wire with proximity induced superconductivity as a function of control parameters $\Delta'$ and $\Delta_z$ in units of $\Delta_0\neq0$. In the parameter regimes $0\leq\Delta_z<(\Delta_0^2-\Delta'^2)/\Delta_0$ the system is nontopological, while when $\Delta_z>|\Delta_0^2-\Delta'^2|/\Delta_0$ and $0<\Delta_z<(\Delta'^2-\Delta_0^2)/\Delta_0$ the system supports one and two pairs of MFs, respectively. When $\Delta_z=0$ and $\Delta'>\Delta_0$, although two MF solutions can be found, the system is gapless and thus the MFs are not localized states.}
\label{phase2}
\end{figure}

Our results are summarized as a phase diagram (Fig.~\ref{phase2}) as a function of $\Delta'$ and $\Delta_z$, fixing $\Delta_0\neq0$. For a sufficiently small magnetic field and decrease in superconductivity, the system is a conventional $s$-wave superconductor and therefore has no MFs. When $\Delta_0\Delta_z$ is larger than the difference in squares of $\Delta_0$ and $\Delta'$, the system is in the known phase of the one-dimensional topological superconductor supporting a single pair of MFs.\cite{klinovajaPRB12} Finally, when $\Delta'$ is larger than the superconducting gap in the bulk and the magnetic exchange energy is sufficiently small but finite,  $0<\Delta_z<(\Delta'^2-\Delta_0^2)/\Delta_0$, we find the system is capable of supporting two pairs of MFs, with two MFs on the left end of the system and two on the right end.

\section{Conclusions}
We have studied a chain of helically ordered magnetic impurities on a conventional $s$-wave superconductor, including a periodic spatial decrease in the proximity induced superconducting order parameter: first, when the band is formed by adjacent Shiba states within the bulk superconductor (Sec.~\ref{shiba}) and second, when the band is formed within the one-dimensional atomic chain with induced superconductivity (Sec.~\ref{qw}). In both cases, if the magnetic exchange energy and the decrease in superconductivity are small compared to the bulk superconducting gap, then the systems are topologically trivial. Also for both scenarios, when the magnetic exchange energy is much larger than both the bulk superconductivity and local decrease in superconductivity, the system is able to support one pair of MFs. However, when the local suppression of the superconducting order parameter is large, comparable to the bulk order parameter itself, then the two models differ: for the Shiba band there are no MFs while in the quantum wire regime there are two pairs of MFs. 

For many experimental setups where magnetic adatoms are placed on the surface\cite{nadj-pergeSCI14,pawlakCM15}, the magnetic exchange interaction cannot generate a large enough decrease in superconducting order to move away from the single MF pair topological phase.\cite{rusinovJETP69,schlottmanPRB76,salkolaPRB97,balatskyRMP06,flattePRL97,flattePRB97,flatteSSP99,mengPRB15,hoffmanPRB15,bjornsonCM15} Alternatively, we propose an experimental setup to observe two pairs of MFs wherein a quantum wire is placed on top of bulk superconductor with spatially alternating $0$-$\pi$ phase along the direction of the wire.\cite{buzdinRMP05,ryazanovPRL01,kontosPRL02} Perhaps the simplest realization of such a system is to make a two-dimensional superconductor-ferromagnet-superconductor junction with periodically varying thickness of the ferromagnetic layer. By grounding one superconductor, the other will have a spatially alternating $0$-$\pi$ phase according to the period of change in thickness of the ferromagnetic layer. A quantum wire on top of such a superconductor inherits this phase, due to proximity induced superconductivity. Because, for our purpose, a $\pi$ phase is effectively a decrease in the superconducting order by twice the the bulk value, we expect that in such an experimental setup, for sufficiently small magnetic exchange interaction, one may detect two MFs at a single end of the wire.

\begin{acknowledgments}
We acknowledge support from the Swiss NSF and NCCR QSIT. 
\end{acknowledgments}

%\bibliography{mybib}


\begin{appendix}
\begin{widetext}
\section{Shiba Eigenvectors Identities}
\label{idS}
The solutions to Eq.~(\ref{BdG}) satisfy a number of identities which, for completeness we list here,
\begin{align}
&\textbf s_j\cdot\boldsymbol\sigma|\psi_\sigma/\chi_\sigma,j\rangle=\sigma|\psi_\sigma/\chi_\sigma,j\rangle\,,\nonumber\\
&i\tau_y|\psi_\sigma,j\rangle=-|\chi_\sigma,j\rangle\,,\,\,\,i\tau_y|\chi_\sigma,j\rangle=|\psi_\sigma,j\rangle\,,\nonumber\\
&\tau_z|\psi_\sigma,j\rangle=-|\chi_\sigma,j\rangle\,,\,\,\,\tau_z|\chi_\sigma,j\rangle=-|\psi_\sigma,j\rangle\,,
\end{align}
and
\begin{align}
\langle\psi_+,i|\psi_+,j\rangle&=\langle\uparrow,i|\uparrow,j\rangle=\cos(\theta_i/2)\cos(\theta_j/2)+e^{i\phi_j-i\phi_i}\sin(\theta_i/2)\sin(\theta_j/2)\,,\nonumber\\
\langle\psi_-,i|\psi_-,j\rangle&=\langle\downarrow,i|\downarrow,j\rangle=e^{i\phi_i-i\phi_j}\sin(\theta_i/2)\sin(\theta_j/2)+\cos(\theta_i/2)\cos(\theta_j/2)\,,\nonumber\\
\langle\psi_+,i|\psi_-,j\rangle&=-\langle\uparrow,i|\downarrow,j\rangle=-e^{-i\phi_i}\sin(\theta_i/2)\cos(\theta_j/2)+e^{-i\phi_j}\sin(\theta_j/2)\cos(\theta_i/2)\,,\nonumber\\
\langle\psi_-,i|\psi_+,j\rangle&=-\langle\downarrow,i|\uparrow,j\rangle=-e^{i\phi_j}\sin(\theta_j/2)\cos(\theta_i/2)+e^{i\phi_i}\sin(\theta_i/2)\cos(\theta_j/2)\,,\nonumber\\
\langle\psi_{\sigma},i|\chi_{\sigma'},j\rangle&=0\,.
\end{align}

We perform a unitary rotation\cite{pientkaPRB13} to obtain the more convenient form
\begin{align}
&\langle\uparrow,i|\uparrow,j\rangle=e^{(i\phi_i-i\phi_j)/2}\cos(\theta_i/2)\cos(\theta_j/2)+e^{-(i\phi_i-i\phi_j)/2}\sin(\theta_i/2)\sin(\theta_j/2)\,,\nonumber\\
&\langle\downarrow,i|\downarrow,j\rangle=e^{(i\phi_i-i\phi_j)/2}\sin(\theta_i/2)\sin(\theta_j/2)+e^{-(i\phi_i-i\phi_j)/2}\cos(\theta_i/2)\cos(\theta_j/2)\,,\nonumber\\
&\langle\uparrow,i|\downarrow,j\rangle=-e^{-(i\phi_i-i\phi_j)/2}\sin(\theta_i/2)\cos(\theta_j/2)+e^{(i\phi_i-i\phi_j)/2}\sin(\theta_j/2)\cos(\theta_i/2)\,,\nonumber\\
&\langle\downarrow,i|\uparrow,j\rangle=-e^{-(i\phi_i-i\phi_j)/2}\sin(\theta_j/2)\cos(\theta_i/2)+e^{(i\phi_i-i\phi_j)/2}\sin(\theta_i/2)\cos(\theta_j/2)\,,
\end{align}
which is, written in terms of the helix wave vector, $k_h$,
\begin{align}
&\langle\uparrow,i|\uparrow,j\rangle=e^{ik_h x_{ij}}\cos^2(\theta/2)+e^{-ik_h x_{ij}}\sin^2(\theta/2)\,,\nonumber\\
&\langle\downarrow,i|\downarrow,j\rangle=e^{ik_h x_{ij}}\sin^2(\theta/2)+e^{-ik_h x_{ij}}\cos^2(\theta/2)\,,\nonumber\\
&\langle\uparrow,i|\downarrow,j\rangle=-e^{-ik_h x_{ij}}\sin\theta+e^{ik_h x_{ij}}\sin\theta=2i\sin\theta\sin k_h x_{ij}\,,\nonumber\\
&\langle\downarrow,i|\uparrow,j\rangle=-e^{-ik_h x_{ij}}\sin\theta+e^{ik_h x_{ij}}\sin\theta=2i\sin\theta\sin k_h x_{ij}\,,
\end{align}
where $x_{ij}=(i-j)a$.

\section{Fourier Transform}
\label{aft}
We would like to Fourier transform the spatially dependent quantities found in Sec.~\ref{shiba}. For a planar helical spin texture, the entries in Eq.~(\ref{heff1}) and Eq.~(\ref{Heff}) can, in general be written as
\eq{
\frac{e^{-r_{ij}/\xi_\epsilon}}{k_F r_{ij}}A_{\sigma}\left(e^{i k_F r_{ij}}+\sigma e^{-i k_F r_{ij}}\right)\left(e^{i k_h x_{ij}}+\sigma' e^{-i k_h x_{ij}}\right)\,,
}
using the identities from the Appendix~\ref{idS}. Performing a Fourier transform, we find
\begin{align}	
&\sum^{\infty}_{j=-\infty}\frac{e^{-r_{ij}/\xi_\epsilon}}{k_F r_{ij}}e^{i k x_{ij}}A_\sigma\left(e^{i k_F r_{ij}}+\sigma e^{-i k_F r_{ij}}\right)\left(e^{i k_h x_{ij}}+\sigma' e^{-i k_h x_{ij}}\right)\nonumber\\
=&\sum^{\infty}_{j=1}\frac{e^{-a j/\xi_\epsilon}}{k_F a j}A_\sigma\left(e^{i k_F aj}+\sigma e^{-i k_F aj}\right)\left[e^{i k aj}\left(e^{i k_h aj}+\sigma' e^{-i k_h aj}\right)+e^{-i k aj}\left(e^{-i k_h aj}+\sigma' e^{i k_h aj}\right)\right]\nonumber\\
=&\sum^{\infty}_{j=1}\frac{e^{-a j/\xi_\epsilon}}{k_F a j}A_\sigma\left(e^{i k_F aj}+\sigma e^{-i k_F aj}\right)\left[(e^{i k_+ a j}+e^{-i k_+ a j})+\sigma'(e^{i k_- a j}+e^{-i k_- a j})\right]\nonumber\\
=&\sum^{\infty}_{j=1}\frac{e^{-a j/\xi_\epsilon}}{k_F a j}\Re/\Im\left\{\left[e^{i (k_F +k_+) aj}+e^{i (k_F -k_+) aj}\right]+\sigma'\left[e^{i (k_F +k_-) aj}+e^{i (k_F -k_-) aj}\right]\right\}\nonumber\\
=&-\frac{1}{k_F a}\Re/\Im\ln\left[f(k_F+k_+)f(k_F-k_+)f(k_F+k_-)^{\sigma'}f(k_F-k_-)^{\sigma'}\right]\,,
\end{align}
where $A_\sigma=1/2i^{(1-\sigma)/2}$, $k_\pm=k\pm k_h$,  $\Re[F(k_F,k_\pm)] = [F(k_F,k_\pm)+F(-k_F,-k_\pm)]/2$ \{$\Im[F(k_F,k_\pm) ]= [F(k_F,k_\pm)-F(-k_F,-k_\pm)]/2i$\} corresponds to $\sigma=1$ ($\sigma=-1$), $f(k)=1-\exp(-a/\xi_\epsilon+i k a)$, and we have used the summation
\eq{
\sum_{j=1}^\infty \frac{x^j}{j}=-\ln(1-x)\,,
}
for $|x|< 1$.

\section{$\alpha'\gg\alpha$ Shiba Band}
\label{asb}
Projecting Eq.~(\ref{EoM1}) onto the four bands, we obtain a $4N\times4N$ matrix equation, $H^{\textrm{eff}}_2\phi_2=E\phi_2$ with
\begin{equation}
H^{\textrm{eff}}_2=\left(\begin{array}{cc}
\bar h^{\textrm{eff}}_{ij} & \bar\Delta^{\textrm{eff}}_{ij}\\
 \bar\Delta^{\textrm{eff}}_{ij} & -\bar h^{\textrm{eff}}_{ij}
\end{array}\right)
\label{Heff}
\end{equation}
in the basis $(\psi_{+i},\psi_{-i},\chi_{-i},\chi_{+i})$, where (see Appendix~\ref{idS})
\eq{
\bar h^\textrm{eff}_{ii}=\Delta_0\left(\begin{array}{cc}
1-\alpha-\alpha' & 0\\
0 & 1+\alpha-\alpha'
\end{array}\right)\,,\,\,\,\,\,\,\,\,\,\bar \Delta^{\textrm{eff}}_{ii}=0\, ,
}
and 
\begin{align}
\bar h^{\textrm{eff}}_{ij}&=\Delta_0\frac{e^{-r_{ij}/\xi_\epsilon}}{k_F r_{ij}}\sin (k_F r_{ij})\left(\begin{array}{cc}
-\langle\uparrow,i|\uparrow,j\rangle & \langle\uparrow,i|\downarrow,j\rangle\\
\langle\downarrow,i|\uparrow,j\rangle & -\langle\downarrow,i|\downarrow,j\rangle
\end{array}\right)\,,\nonumber \\
\bar \Delta^{\textrm{eff}}_{ij}&=\Delta_0\frac{e^{-r_{ij}/\xi_\epsilon}}{k_F r_{ij}}\cos (k_F r_{ij})\left(\begin{array}{cc}
\langle\uparrow,i|\downarrow,j\rangle & -\langle\uparrow,i|\uparrow,j\rangle\\
-\langle\downarrow,i|\downarrow,j\rangle & \langle\downarrow,i|\uparrow,j\rangle
\end{array}\right)\,,\nonumber \\
\end{align}
for $i\neq j$.
Upon Fourier transforming (see Appendix~\ref{aft}), we obtain a $4\times4$ matrix
\eq{
\mathcal  H_2^{\textrm{eff}}=\tau_z\otimes(h_1 \mathbb1+h_2\sigma_x+\Delta_z\sigma_z)+\tau_x\otimes(\Delta_1 \mathbb1+\Delta_2\sigma_x)\,,
}
where
\begin{align}
h_2&=-\frac{\Delta_0}{k_F a}\Im\left\{\ln\left[\frac{f(k_F+k_+)f(k_F-k_+)}{f(k_F+k_-)f(k_F-k_-)}\right]\right\}\,,\nonumber\\
\Delta_2&=\frac{\Delta_0}{k_F a}\Re\left\{\ln\left[f(k_F+k_+)f(k_F-k_+) f(k_F+k_-)f(k_F-k_-)\right]\right\}\, ,\nonumber\\
\Delta_z&=-\alpha\Delta_0\,.
\label{melems}
\end{align}
$\Delta_1$ is as in Eq.~(\ref{ham12}) while $h_1$ is as in Eq.~(\ref{ham12}) after taking $\alpha=0$. The eigenvalues of this matrix are
\begin{align}
\varepsilon_\pm^2=&h_1^2+h_2^2+\Delta_1^2+\Delta_2^2+\Delta_z^2\pm2\sqrt{(h_1h_2+\Delta_1\Delta_2)^2+\Delta_z^2(h_1^2+\Delta_2^2)}\,.
\end{align}
Because $\alpha\ll1$, we take $\Delta_z=0$, wherein the zero energy eigenvectors are $(0,~0,~\pm1,~1) e^{-i k_\nu^\pm r-\kappa^\pm_\nu r}$, where $\nu=5,6$, when $1/a\gg k_F,~k_h$ and the condition on $k$, so that $\varepsilon_\pm=0$, simplifies to $h_1-i \Delta_1=\mp(h_2-i \Delta_2)$. 
In the small spacing limit, we find four Dirac points at 
\begin{align}
k^\pm_5  = - k^\mp_6= \pm k_h + \pi/3a
\end{align}
with decays
\begin{equation}
\kappa_5^\pm=-\kappa^\pm_6=k_F\frac{2+(1-\alpha')}{2\sqrt{3}}\,.
\label{wv}
\end{equation}
Applying the same argument as in the two band model, because $\kappa_5^\pm\kappa_6^\pm<0$, we cannot construct a MF that satisfies the boundary conditions.

\section{Quantum Wire Zero Energy Wave Functions}
\label{aqw}
The full  $8\times8$ Hamiltonian density, $\mathcal H=\mathcal H^0+\mathcal H^I$, from Sec.~\ref{qw} is, in matrix form,
\eq{
\left(
\begin{array}{cccccccc}
-i\hbar v_F\partial_r & \Delta_z & 0 & -\Delta_s & 0 & 0 & 0 & -e^{i\theta}\Delta'\\
\Delta_z & i\hbar v_F \partial_r & \Delta_s & 0 & 0 &0 & e^{-i\theta}\Delta' & 0 \\
0 & \Delta_s  & -i\hbar v_F \partial_r & -\Delta_z & 0 & e^{-i\theta}\Delta' & 0 & 0 \\
-\Delta_s&0&-\Delta_z&i\hbar v_F\partial_r& -e^{i\theta}\Delta' & 0 & 0 &0 \\
0 & 0 & 0 &  -e^{-i\theta}\Delta' & i\hbar v_F \partial_r & 0 & 0 & -\Delta_s \\
0 & 0 &  e^{i\theta}\Delta' & 0 & 0 & -i\hbar v_F \partial_r & \Delta_s & 0 \\
0 &  e^{i\theta}\Delta' & 0 & 0 & 0 & \Delta_s & i\hbar v_F \partial_r & 0 \\
 -e^{-i\theta}\Delta' & 0 & 0 & 0 & -\Delta_s & 0 & 0 & -i\hbar v_F \partial_r\\
\end{array}
\right)
}
with four zero energy solutions
\eq{
\phi_1=\left( \begin{array}{c} i\frac{\Delta_0-\hbar v_F\kappa^\pm_1}{\Delta'} \\ -\frac{\Delta_0-\hbar v_F\kappa^\pm_1}{\Delta'}  \\ -i\frac{\Delta_0-\hbar v_F\kappa^\pm_1}{\Delta'} \\ -\frac{\Delta_0-\hbar v_F\kappa^\pm_1}{\Delta'} \\-i\\1\\i\\1\end{array} \right) e^{\kappa^\pm_1 r}~,~~
\phi_2=\left( \begin{array}{c} i\frac{\Delta_0-\hbar v_F\kappa^\pm_2}{\Delta'} \\ \frac{\Delta_0-\hbar v_F\kappa^\pm_2}{\Delta'}  \\ -i\frac{\Delta_0-\hbar v_F\kappa^\pm_2}{\Delta'} \\ \frac{\Delta_0-\hbar v_F\kappa^\pm_2}{\Delta'} \\i\\1\\-i\\1\end{array} \right)e^{\kappa^\pm_2 r}~,~~
\phi_3=\left( \begin{array}{c} -i\frac{\Delta_0+\hbar v_F\kappa^\pm_3}{\Delta'} \\ -\frac{\Delta_0+\hbar v_F\kappa^\pm_3}{\Delta'}  \\ i\frac{\Delta_0+\hbar v_F\kappa^\pm_3}{\Delta'} \\ -\frac{\Delta_0+\hbar v_F\kappa^\pm_3}{\Delta'} \\i\\1\\-i\\1\end{array} \right)e^{\kappa^\pm_3 r}~,~~
\phi_4=\left( \begin{array}{c} -i\frac{\Delta_0+\hbar v_F\kappa^\pm_4}{\Delta'} \\ \frac{\Delta_0+\hbar v_F\kappa^\pm_4}{\Delta'}  \\ i\frac{\Delta_0+\hbar v_F\kappa^\pm_4}{\Delta'} \\ \frac{\Delta_0+\hbar v_F\kappa^\pm_4}{\Delta'} \\-i\\1\\i\\1\end{array} \right)e^{\kappa^\pm_4 r}~,~~
\label{sols}
}
in the basis $(R_\uparrow,L_\downarrow,R_\uparrow^\dagger,L_\downarrow^\dagger,L_\uparrow,R_\downarrow,L_\uparrow^\dagger,R_\downarrow^\dagger)$. Upon 

Transforming from the basis of the left-right moving operators to the electron-hole operators $(\Psi_\uparrow,\Psi_\downarrow,\Psi^\dagger_\uparrow,\Psi^\dagger_\downarrow)$ and dropping fast oscillating terms, we find the solutions given in
Eq.~(\ref{sols}) to become
\begin{align}
\Phi^\pm_1&=\left[-\frac{\hbar v_F\kappa^\pm_1+\Delta_s}{\Delta'}\left( \begin{array}{c} -ie^{-i k_F r}\\e^{i k_F r}\\ie^{i k_F r}\\e^{-i k_F r}\end{array} \right)+\left( \begin{array}{c} -ie^{i k_F r}\\e^{-i k_F r}\\ie^{-i k_F r}\\e^{i k_F r}\end{array} \right)\right] e^{-\kappa^\pm_1 r}\,,\nonumber\\
\Phi^\pm_2&=\left[-\frac{\hbar v_F\kappa^\pm_2+\Delta_s}{\Delta'}\left( \begin{array}{c} e^{-i k_F r}\\-ie^{i k_F r}\\e^{i k_F r}\\ie^{-i k_F r}\end{array} \right)+\left( \begin{array}{c} e^{i k_F r}\\-ie^{-i k_F r}\\e^{-i k_F r}\\ie^{i k_F r}\end{array} \right)\right] e^{-\kappa^\pm_2 r }\,,
\nonumber\\
\Phi^\pm_3&=\left[\frac{-\hbar v_F\kappa^\pm_3+\Delta_s}{\Delta'}\left( \begin{array}{c} -ie^{-i k_Fr}\\-e^{i k_Fr}\\ie^{i k_Fr}\\-e^{-i k_Fr}\end{array} \right)-\left( \begin{array}{c} -ie^{i k_Fr}\\-e^{-i k_Fr}\\ie^{-i k_Fr}\\-e^{i k_Fr}\end{array} \right)\right] e^{-\kappa^\pm_3 r}\,,\nonumber\\
\Phi^\pm_4&=\left[\frac{-\hbar v_F\kappa^\pm_4+\Delta_s}{\Delta'}\left( \begin{array}{c} e^{-i k_Fr}\\ie^{i k_Fr}\\e^{i k_Fr}\\-ie^{-i k_Fr}\end{array} \right) -\left( \begin{array}{c} e^{i k_Fr}\\ie^{-i k_Fr}\\e^{-i k_Fr}\\-ie^{i k_Fr}\end{array} \right)\right]e^{-\kappa^\pm_4 r}\,,
\end{align}
respectively.
\end{widetext}

\end{appendix}

\end{document}